\documentclass[twocolumn,aps,prl,showpacs,superscriptaddress]{revtex4-1}
\usepackage[latin9]{inputenc}
\usepackage{amsmath}
\usepackage{amssymb}
\usepackage{graphicx}
\usepackage{ulem}
\usepackage[colorlinks,linkcolor=blue,anchorcolor=blue,citecolor=blue,urlcolor=blue]{hyperref}
\makeatletter
\usepackage{color}
\usepackage{bm}
\usepackage{float}

\makeatother

\begin{document}

\title{Cross-Filament Stochastic Acceleration of Electrons in Kilojoule Picosecond Laser \\Interactions with Near Critical Density Plasmas}

\author{X. F. Shen}

\affiliation{Institut f\"ur Theoretische Physik I, Heinrich-Heine-Universit\"at D\"usseldorf,
40225 D\"usseldorf, Germany}

\author{A. Pukhov}
\email{pukhov@tp1.uni-duesseldorf.de}

%\selectlanguage{english}%
%\selectlanguage{english}%

\affiliation{Institut f\"ur Theoretische Physik I, Heinrich-Heine-Universit\"at D\"usseldorf,
40225 D\"usseldorf, Germany}

\author{O. N. Rosmej}
\affiliation{GSI Helmholtzzentrum f\"ur Schwerionenforschung GmbH, Planckstr.1, 64291 Darmstadt, Germany}
\affiliation{Helmholtz Forschungsakademie Hessen f\"ur FAIR (HFHF), Campus Frankfurt am Main, Max-von-Laue-Stra\ss e 12,
60438 Frankfurt am Main, Germany}

\author{N. E. Andreev}
\affiliation{Joint Institute for High Temperatures, RAS, Izhorskaya st.13, Bldg. 2, 125412, Moscow, Russia}
\affiliation{Moscow Institute of Physics and Technology (State University), Institutskiy Pereulok 9, 141700, Dolgoprudny Moscow Region, Russia}

\date{\today}

\begin{abstract}
	Understanding the interaction of kilojoule, picosecond laser pulse with long-scale length preplasma or homogeneous near critical density (NCD) plasma 
	is crucial for guiding experiments at national short-pulse laser facilities. 
Using full three-dimensional particle-in-cell simulations, we demonstrate that in this regime, cross-filament stochastic acceleration is an important mechanism that contributes to the production of superponderomotive, high-flux electron beams. Since the laser power significantly exceeds the threshold of the relativistic self-focusing, multiple filaments are generated and can propagate independently over a long distance. Electrons jump across the filaments during the acceleration, and their motion becomes stochastic.  
We find that the effective temperature of electrons increases with the total interaction time following a scaling like $T_{\rm eff}\propto\tau_{i}^{0.65}$.   
By irradiating a submillimeter thick NCD target,  
the space charge of electrons with energy above 2.5 MeV reaches tens of $\mu$C.
Such high-flux electrons with superponderomotive energies significantly facilitate  applications in high-energy-density science, nuclear science, secondary sources and diagnostic techniques. 
\end{abstract}

\maketitle

\section{INTRODUCTION}

The advent of kilojoule laser facilities, like NIF-ARC \cite{Crane2010}, LMJ-PETAL \cite{Batani2014}, and LFEX \cite{Miyanaga2006}, opens a new frontier of laser-plasma interaction (LPI). It has attracted great attention  \cite{Li2008,Kemp2012,Sorokovikova2016,Ferri2016,Chen2017,Yogo2017,Iwata2018,Kim2018,Iwata2018,Mariscal2019,Kemp2020,Matsuo2020,Williams2020,Hussein2021,Simpson2021,Raffestin2021} due to its broad  applications in high-energy-density (HED) science \cite{Matsuo2020,Drake2018}, probing \cite{Chen2017,Simpson2021b} and driving \cite{Robinson2014,Roth2001} inertial confinement fusion (ICF),  production of high-flux electron beams \cite{Kemp2012,Sorokovikova2016,Hussein2021} and secondary sources (e.g., ions, neutrons, x/$\gamma$-rays) \cite{Yogo2017,Simpson2021,Raffestin2021}, laboratory astrophysics \cite{Chen2015}, nuclear science \cite{Mizutania2020}, etc. These lasers are limited to deliver picosecond pulses with large focal spots of tens of micrometers due to both technological and infrastructural constraints \cite{Simpson2021}. The interaction physics of such pulses with plasmas is quite different from the usually discussed subpicosecond pulses with diffraction-limited small focal spots (typically $<$ 10$\mu$m) \cite{Wilks1992,Gibbon1996,Zepf2003,Wei2004,Robinson2011,Rosmej2020,Shen2021,Shen2021a}. On the one hand, the long interaction time brings us to the mesoscale between kinetic and fluid regimes; on the other hand, the large focal spot results in quasi-onedimensional plasma expansion in laser-solid target interactions.  
However, in underdense plasmas, it may trigger filamentation instability  \cite{Pukhov1996,Tanaka2000,Pukhov2003,Macchi2007} since the laser power is several orders of magnitude higher than the threshold of relativistic self-focusing \cite{Max1974,Borisov1992}. 
Experiments and simulations have observed new results, specifically the enhancement of scalings of electron \cite{Sorokovikova2016,Mariscal2019,Kemp2020,Williams2020,Simpson2021} and proton acceleration \cite{Yogo2017,Simpson2021,Raffestin2021}, but the debate about the acceleration mechanisms still remains controversial.

Previous studies of such laser pulses focused on the interaction with overdense plasmas and effects of the long pulse duration.  
Theoretical investigations mainly relied on one- or two-dimensional (2D) particle-in-cell (PIC) simulations, since three-dimensional (3D) simulations demand large computational resources. However, to thoroughly understand the underlying physics and quantitatively explain the experimental results, it is essential to conduct 3D simulation studies. On the other hand, recent experimental and numerical researches have demonstrated that in the interactions of femtosecond or subpicosecond laser pulses with small focal spots with near critical density (NCD) plasmas, the maximum energy and flux of both electrons and ions can be much higher than those obtained with solid targets \cite{Rosmej2019,Rosmej2020,Bin2015,Lobok2018,Willingale2018,Ma2019,Pazzaglia2020}. This is because a long plasma channel is formed due to the self-focusing and coalescence of adjacent filaments, 
in which 
abundant electrons are accelerated dominantly by direct laser acceleration (DLA) \cite{Pukhov1999,Lehe2014,Arefiev2015,Wang2020}. However, when a laser pulse with large focal spot is used, distances between neighbouring filaments may be much larger than the skin depth, so they do not feel each other and the coalescence cannot proceed further \cite{Pukhov2003}. Therefore, many channels survive for a long time. This self-focusing process can only be properly described by 3D simulations, the reason of which was explained in Ref. \cite{Pukhov2003}. Nevertheless, how it affects the LPI on multipicosecond timescale, especially for electron acceleration,   
is still unclear.

\begin{figure*}
	\includegraphics[width=12.cm]{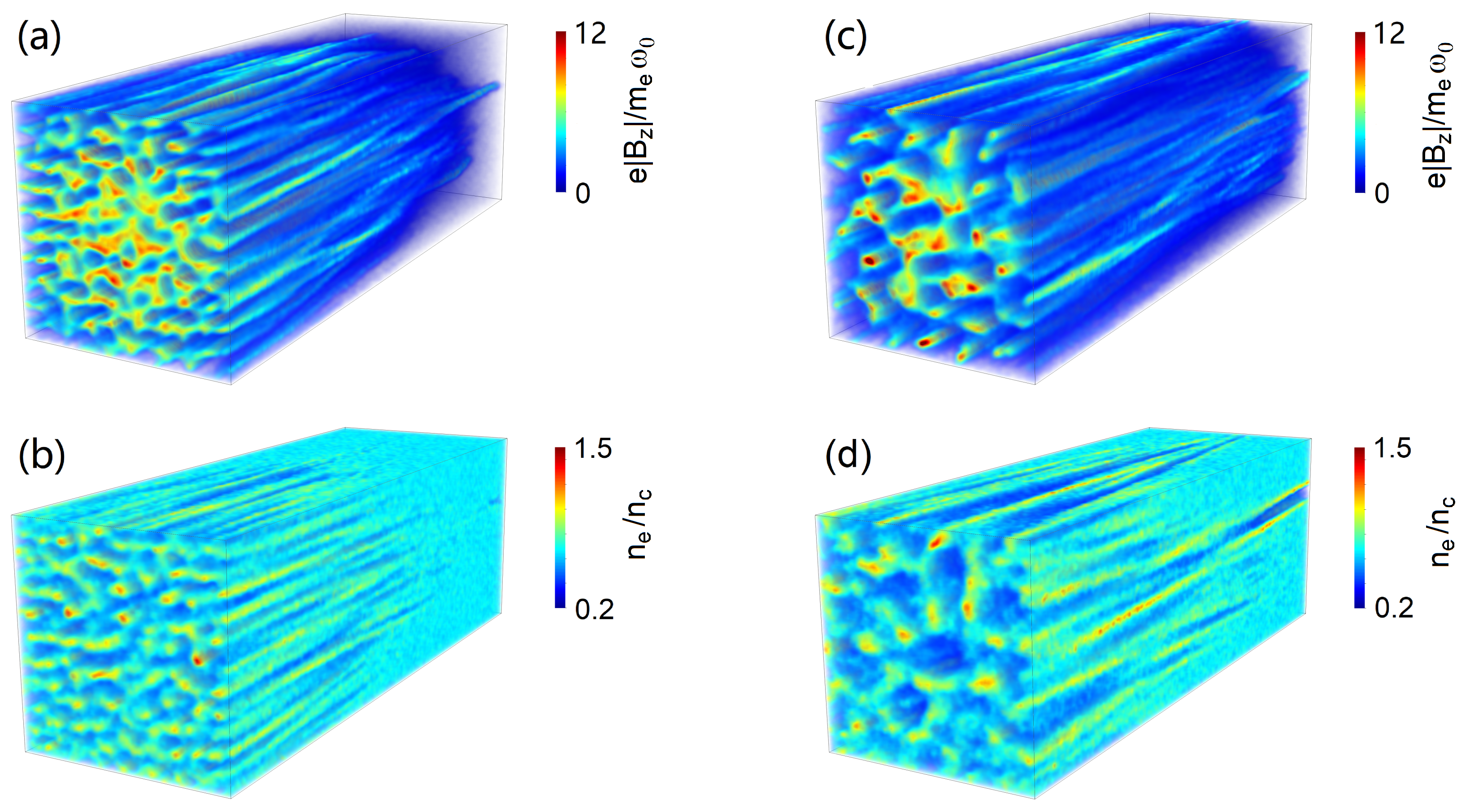}
	\caption{
		\label{fig:fig1}
		3D PIC simulation results of a long NCD plasma irradiated by a kJ, ps laser pulse. The magnetic field $|B_z|$ [(a), (c)] and electron density [(b), (d)] are presented at $t=-0.033$ps [(a), (b)] and $0.367$ps [(c), (d)], respectively, showing that strong filamentation occurs when a large focal-spot laser is used. Here $t=0$ corresponds to the time when the laser peak enters the plasma. Only part of simulation box (within $\pm30\lambda$ and length of 90$\lambda$) is shown.}
\end{figure*}

In this paper, we investigate the long-time interaction of realistic picosecond laser pulses with large focal spots with submillimeter NCD plasmas via full-scale 3D PIC simulations. We find that most energetic electrons are no longer confined in an individual channel. They first are accelerated by DLA in a channel, and then may jump across into adjacent channels. This triggers the stochastic motion of electrons. As a result, electrons can be further heated up to superponderomotive energies.  
The effective electron temperature increases with interaction time (defined by laser pulse duration and depletion in extended NCD plasma) as $T_{\rm eff}\propto\tau_{i}^{0.65}$. 

\section{SIMULATION PARAMETERS AND MAIN RESULTS}

We conducted 3D PIC simulations with the  
VLPL code by considering realistic laser and plasma parameters \cite{VLPL_CERN}. A $y$-polarized laser with intensity of $4.74\times10^{19}$ $\rm W/cm^2$ and wavelength of $\lambda=1\mu$m is incident on a fully-ionized, homogeneous plasma with electron density $n_e=0.65n_{c}$ and length $400\lambda$. Here $n_{c}=m_e\omega_0^2/4\pi e^2$ is the critical density, where $\omega_0$, $m_e$ and $e$ are the laser frequency, electron mass and charge, respectively. The temporal and spatial profiles are Gaussian distributions. The pulse length is $\tau_L=0.7$ps (FWHM) and the whole interaction time reaches over 3ps. The focal spot is $d_L=50\lambda$ (FWHM), leading to a power of 1.5 PW and energy of 1 kJ.  
The ion composition is $n_{\rm C^{6+}}:n_{\rm H^+}:n_{\rm O^{8+}}=3:4:2$ \cite{Pugachev2016,Rosmej2019,Rosmej2020}. The simulation box is $425\lambda\times250\lambda\times250\lambda$. The first $10\lambda$ and last $15\lambda$ space in $x$-direction are vacuum. 
The longitudinal resolution is $h_x=0.1\lambda$. In transverse dimensions, within the focal spot (i.e., $|y,z|<42\lambda$), a finer resolution of $h_y=h_z=0.5\lambda$ is used, while outside this region, the cell size increases exponentially with a factor of 1.05 to save computational resources. A numerical-Cherenkov-free RIP Maxwell solver is used \cite{Pukhov2020}. 
We used 4 macroparticles per cell for electrons and 1 for ions of each type. 
The numerical convergence was confirmed by comparing interested physical quantities at different resolutions.

The laser power is more than four orders of magnitude higher than the power threshold of relativistic self-focusing \cite{Max1974,Borisov1992} $P_{cr}=17(\omega_0/\omega_p)^2\approx26$ GW. Therefore, the laser pulse breaks up into many small filaments when it enters the plasma. Each carries the critical power and undergoes the self-focusing process. 
Figure \ref{fig:fig1} illustrates the temporal evolution of the magnetic fields $|B_z|$ [\ref{fig:fig1}(a), \ref{fig:fig1}(c)] and electron density [\ref{fig:fig1}(b), \ref{fig:fig1}(d)] at $t=-0.033$ps [\ref{fig:fig1}(a), \ref{fig:fig1}(b)] and $0.367$ps [\ref{fig:fig1}(c), \ref{fig:fig1}(d)]. One can clearly see the evidences of the multifilaments in the $B_z$ field and the corresponding multichannels in the density distribution. The channels are not evacuated completely and the residual electron density is about $0.2n_c$. 
During the interaction, the maximum laser intensity reaches $2\times10^{20}\,{\rm W/cm^2}$ due to the self-focusing. 

\begin{figure}
	\includegraphics[width=8.6cm]{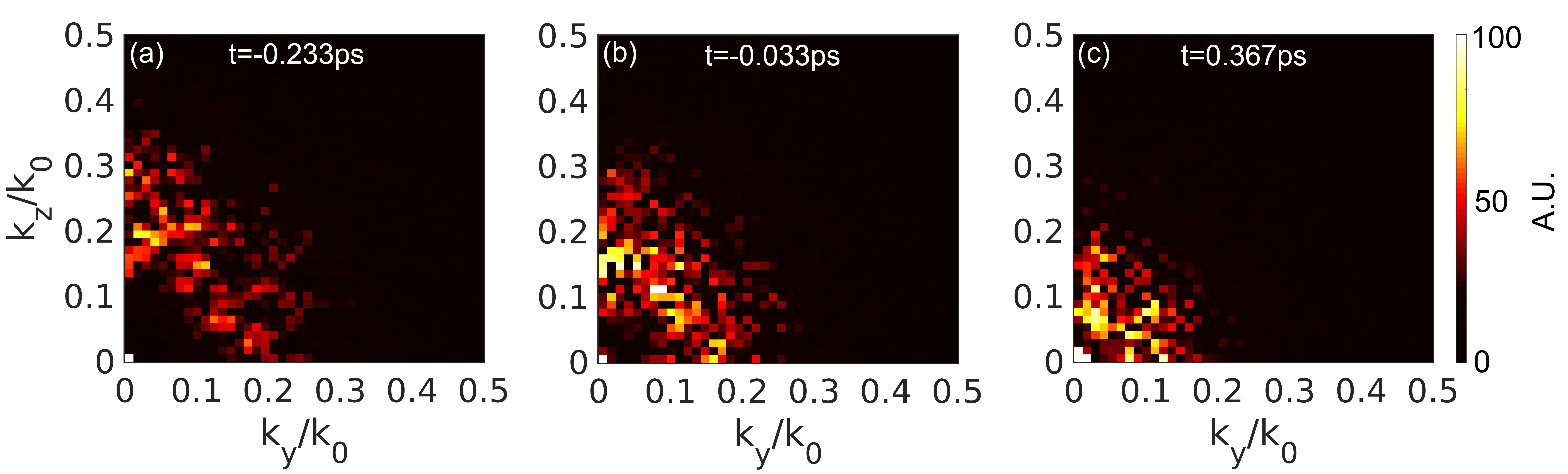} \caption{
		The temporal evolution of the wave number spectra of the $B_z$ field observed at $x=50\lambda$. The color scale corresponds to relative spectral intensity. 	
	}
	\label{fig:fig2} 
\end{figure}

In Fig. \ref{fig:fig2}, we show the temporal evolution of the transverse modes by analyzing the $B_z$ field, where it is evident that during a long time, the dominant mode is around $0.2k_0$, especially when the laser peak enters the plasma ($t=0$). 
This means that the distance between neighbouring filaments is usually much larger than the skin depth $d_s=c/\omega_p$. Therefore, the magnetic fields are shielded by the surrounding plasma, and the coalescence induced by  magnetic attraction cannot further occur naturally \cite{Pukhov2003,Sentoku2000}. The channels can remain almost straight and well separated till $t=0.367$ps. After that, induced by hosing-like instability \cite{Sprangle1994,Huang2017} and transverse expansion of ions \cite{Pukhov2003}, the distances of neighboring channels can become smaller, and some of them merge into several larger ones [Figs. \ref{fig:fig1}(c) and (d)], corresponding to those longer modes in Fig. \ref{fig:fig2}(c). The laser pulse is completely absorbed at $x=375\lambda$ and $t=1.767$ps (about 3 ps since the start of LPI), while the self-generated quasistatic magnetic field can last for even longer time. Note that since the averaged distance between two neighboring filaments is about ten times larger than our transverse resolution, the fundamental physical phenomena should be similar with even higher resolutions.

In our regime, most of the laser energy is absorbed by the plasma. 
This is a key advantage compared to laser-solid target interaction \cite{Gibbon2005,Macchi2013,Shen2017}. The conversion efficiency from laser to particles reaches $84\%$. At $t=2.1$ps, the energy carried by electrons is about 770 J and the rest mainly absorbed by ions, in which protons, carbon ions and oxygen ions carry 16.7 J, 27.1 J and 24.1 J, respectively.    
In VLPL,  
we register all electrons that leave from the simulation box boundaries. The space charge of forward-moving electrons with energy above 2.5 MeV reaches 38 $\mu$C ($\sim 2.4\times10^{14}$), which is more than one order of magnitude higher than that reported in Ref. \cite{Raffestin2021} where solid targets and 450 J laser energy were used.

\begin{figure}[b]
	\includegraphics[width=8.6cm]{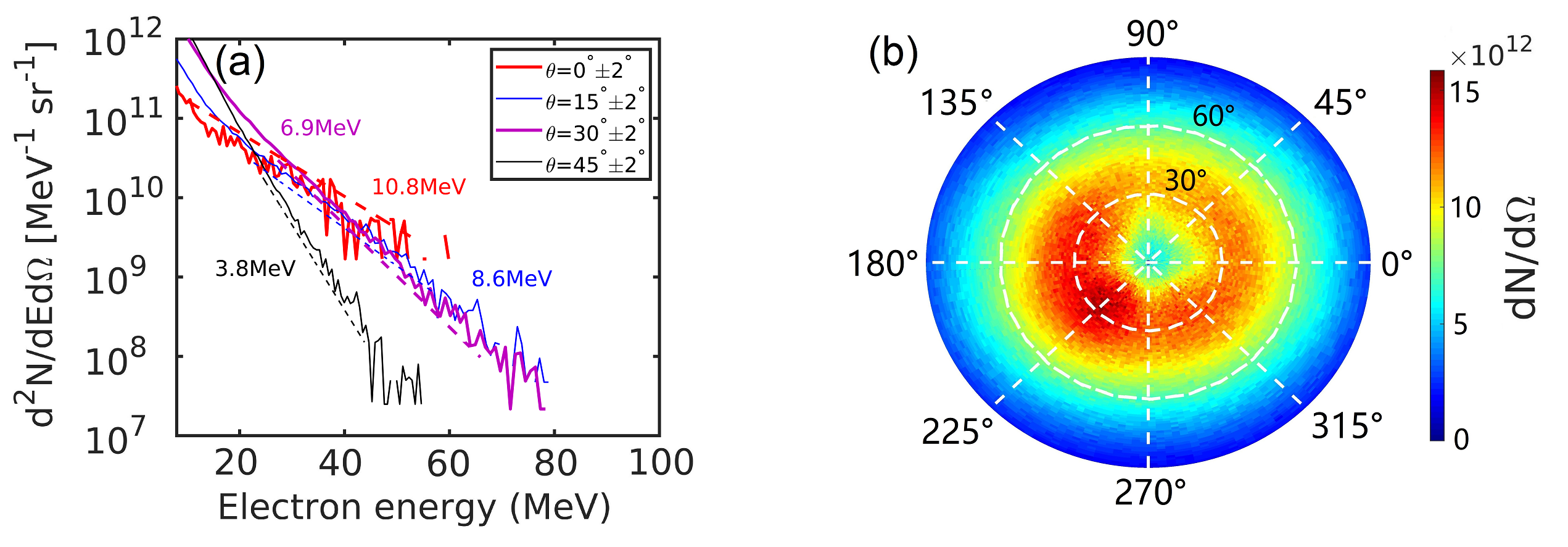}
	\caption{
		\label{fig:fig3} 
		(a) Energy spectra per steradian $d^2N/dEd\Omega$ of electrons that leave the simulation box till $t=2.767$ps around $\theta=0^\circ$ (red), $15^\circ$ (blue), $30^\circ$ (magenta) and $45^\circ$ (black), where $d\Omega={\rm sin}\theta d\theta d\phi$, $\theta={\rm arctan}(\sqrt{p_y^2+p_z^2})/p_x$ and $\phi={\rm arctan}(p_z/p_y)$. (b) Angular distribution of high-energy electrons.  
	}
\end{figure}

In Fig. \ref{fig:fig3}, we show the energy distribution [\ref{fig:fig3}(a)] and divergence [\ref{fig:fig3}(b)] of electrons, in which for electrons registered at the side boundaries, we only consider those high-energy ones with $\gamma>16$. 
This leads to a total space charge of electrons about 7.6 $\mu$C, while the space charge of electrons confined inside $15^\circ$ (almost all of them are registered at the rear boundary) is only about 0.3 $\mu$C.  
A distinctive feature is that the peak brightness appears around $30^\circ$, instead of the laser direction, the difference between which is about a factor of 2.  
This feature can be used to distinguish the CFSA from other acceleration mechanisms  \cite{Sorokovikova2016,Iwata2018,Kemp2020,Pukhov1999,Sheng2002} in experiment.   
The effective temperatures $T_{\rm eff}$ around angles of  $0^\circ$ and $15^\circ$ are 10.8 MeV and 8.6 MeV, respectively, much larger than the ponderomotive scaling \cite{Wilks1992} $T_{\rm pond}=(\sqrt{1+a_0^2/2}-1)m_ec^2\approx$1.6 MeV. At $45^\circ$, the temperature is much lower, about 3.8 MeV.  
Moreover, the relatively low forward flux can be partly attributed to the Alfv\'en current limit \cite{Alfven1939} $J_A=m_ec^3\beta\gamma/e=17\beta\gamma$ kA. Taking the beam length about 1ps and $\gamma=20$ ($\simeq T_{\rm eff}$), it gives 0.34 $\mu$C. 

\begin{figure}
	\includegraphics[width=8.6cm]{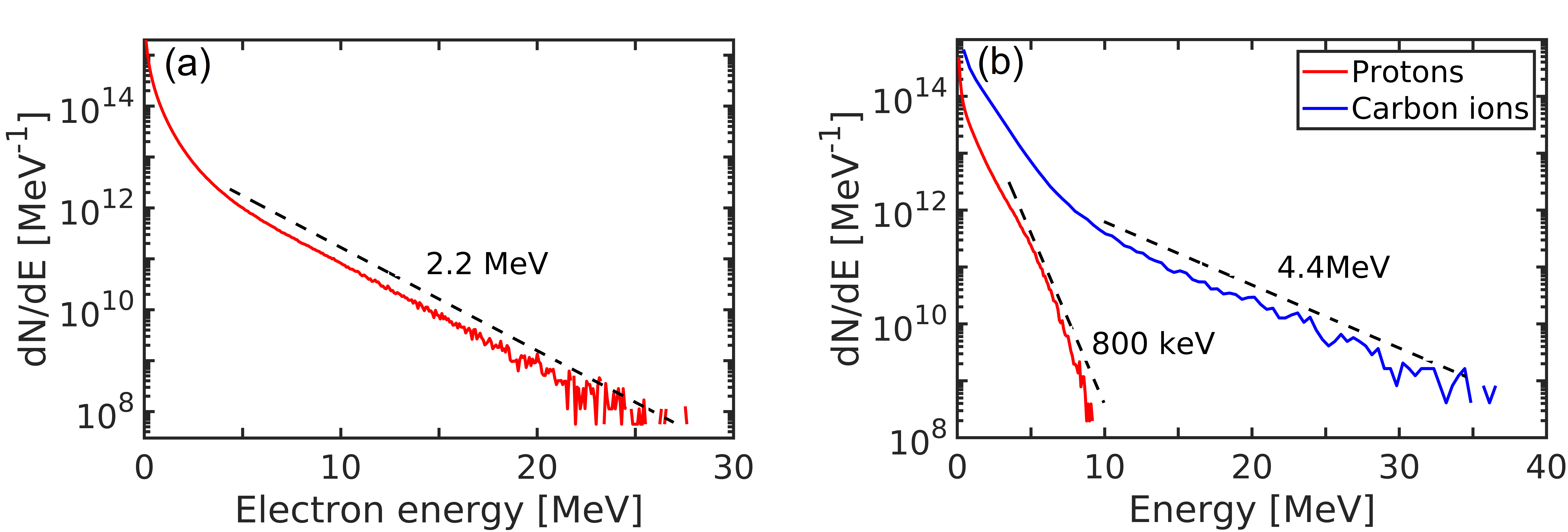} \caption{(a) Energy spectrum of electrons staying inside the simulation box at $t=2.1$ ps. (b) Energy spectra of protons (red) and carbon ions (blue) staying inside the initial plasma boundary (i.e. $10\lambda<x<410\lambda$) at $t=2.1$ ps. 	
	}
	\label{fig:fig4} 
\end{figure}

The effective temperature of electrons inside the box at $t=2.1$ps is about 2.2 MeV, as shown in Fig. \ref{fig:fig4}(a),  
while for protons and carbon ions, their corresponding effective temperatures are about 800 keV and 4.4 MeV (i.e., 367 keV$/\mu$), see the red and blue lines in Fig. \ref{fig:fig4}(b). There are $5\times10^{14}$ protons with energy $>10$ keV and $10^{13}$ protons with energy above the threshold (i.e., 1.64 MeV) of the $\rm ^7Li(p,n)^7Be$ reaction \cite{Brown2018}. Since the characteristic energies of protons and carbon ions (per nucleon) are lower than that of electrons, subsequently, the electron energy will be further transferred to ions via thermal expansion over a much longer time, leading to even more abundant energetic ions. This may find applications in production of high-flux neutrons by replacing part of the plasma components with nuclear materials \cite{Kar2016,Curtis2018,Kemp2019,Jiang2021}.

\begin{figure}
	\includegraphics[width=8cm]{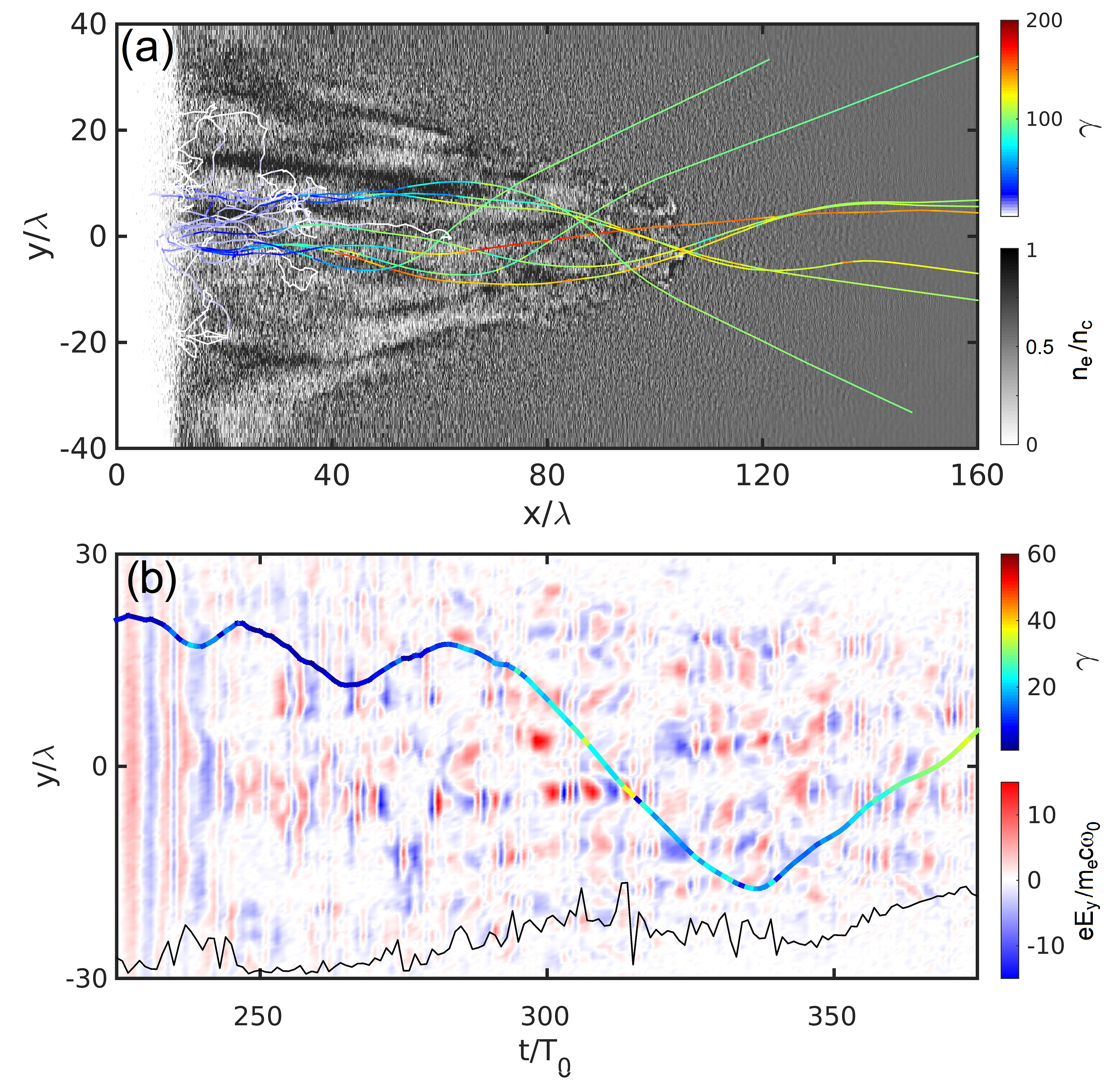}
	\caption{
		\label{fig:fig5} 
		(a) Particle trajectories and density distribution, showing that electrons jump across several filaments during the acceleration. (b) The witnessed $E_y$ field for a representative electron that experiences the CFSA. The black line shows the evolution of $\gamma$-factor. In (a), (b), trajectories of electrons are shown by the lines color-coded with their corresponding $\gamma$-factor.}
\end{figure}

\section{ELECTRON ACCELERATION DYNAMICS}
\subsection{Stochastic acceleration}

To understand the underlying physics, 
we performed 3D simulations with particle tracer. Hereinafter, the laser intensity is fixed at $4.74\times10^{19}\;{\rm W/cm^2}$. A laser pulse with a smaller focal spot of $d_L=35.3\lambda$ and shorter pulse duration of $\tau_L=500$ fs is used to reduce computational cost. 
The plasma density is $0.75n_c$ to deplete the pulse in a shorter distance.  
In Fig. \ref{fig:fig5}(a), we present the 2D cut of the electron density distribution at $t=0.117$ps, where many filaments are formed. The solid lines, color-coded with respect $\gamma$-factor, depict the trajectories of several representative electrons. One can see that these electrons jump across multichannels during their acceleration, different from that in DLA where resonant electrons are trapped in a single channel and undergo betatron oscillation  \cite{Pukhov1999,Rosmej2020}. When the electrons jump from one channel to the neighboring channels, their motions become stochastic. This is because in different channels, the phase of the laser field, determined by the residual electron density and the channel direction, is different. Here the origin of the stochastic motion is different from that in the well-known regime, where it is induced by two counterpropagating pulses \cite{Sheng2002,Sheng2004}. 

Figure \ref{fig:fig5}(b) shows the cross-filament stochastic acceleration (CFSA) process of a representative electron, where the electron trajectory (color line) is plotted on top of the transverse field in a comoving window with width of $\Delta_y=\pm30\lambda$. The black line illustrates the time evolution of the electron $\gamma$. One can see that the electron experiences lots of random kicks (corresponding to oscillations of $\gamma$) as it jumps across the filaments and the value of $\gamma$ is slowly increasing.

\begin{figure}
	\includegraphics[width=8.6cm]{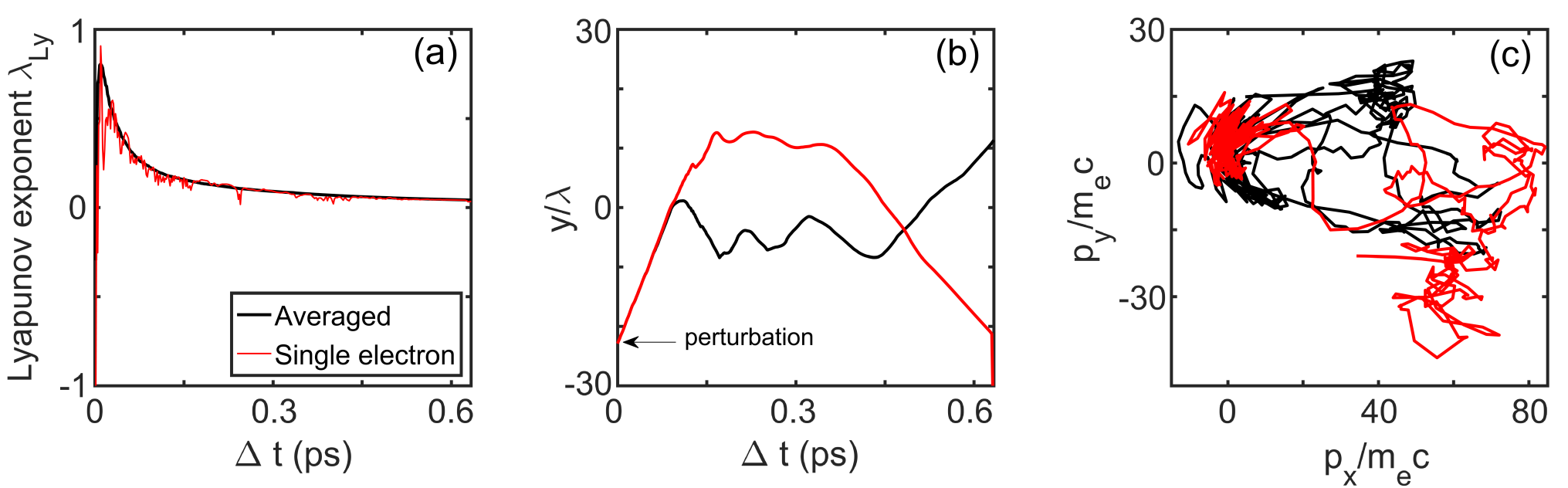}
	\caption{
		\label{fig:fig6}
		 Stochasticity of electron acceleration. (a) Time evolution of the $\lambda_{Ly}$, demonstrating that the electron motion is stochastic. The black line shows the averaged $\lambda_{Ly}$, while the red line is from a single electron. 
		  Trajectories (b) and momenta space (c) of this electron with (red) and without (black) perturbation are shown.}
\end{figure}

To evaluate the stochasticity of the system, we calculate the Lyapunov exponent $\lambda_{Ly}$ \cite{Lichtenberg1984,Sentoku2003}
\begin{eqnarray}
	\lambda_{Ly}=\frac{2\pi}{\omega_0\Delta t}{\rm ln}\frac{\sum|\bm{P(P_0}+\delta \bm{ P)-P(P_0)}|}{\sum|\delta {\bm P}|},\label{eq:Ly}
\end{eqnarray}
where $\bm {P(P_0)}$ and $\bm{P(P_0}+\delta \bm{P)}$ are electron momentum without and with perturbation, respectively, $\delta \bm{P}$ is initial momentum perturbation, and $\Delta t$ is the time passed after the perturbation added in simulation. Here we add $\delta \bm{P}=\delta P_x=0.01m_ec$ at $t=-0.2$ps. For negative $\lambda_{Ly}$, the system is stable to small perturbations. However, if $\lambda_{Ly}$ is positive, 
the system is chaotic. The calculation of Eq. (\ref{eq:Ly}) is shown in Fig. \ref{fig:fig6}(a), where the black line represents the Lyapunov exponent $\lambda_{Ly}$ averaged over 300 particles whose initial positions are randomly distributed near the front of the plasma, and the red line shows the $\lambda_{Ly}$ for a representative single particle. In Figs. \ref{fig:fig6}(b) and \ref{fig:fig6}(c), we present the trajectory and momenta space of this electron, respectively, where the black line represents the case without perturbation and the red with perturbation. It is evident that $\lambda_{Ly}$ is always positive and a tiny perturbation results in exponentially diverging trajectories [\ref{fig:fig6}(b)] and momentum space [\ref{fig:fig6}(c)]. Therefore, we conclude that  the electron motion is stochastic and electrons experience the CFSA.

\subsection{Scaling}
Another important feature of the stochastic acceleration is that the effective temperature increases with the interaction time \cite{Meyer-ter-Vehn1999}.  
A similar trend has also been observed in recent experiments, where large focal-spot laser pulses with different $\tau_L$ (determining $\tau_{i}$) are used \cite{Simpson2021}.
In Ref. \cite{Meyer-ter-Vehn1999}, a simple analytical model was proposed to explain this feature, where single electron motion in a planar laser pulse is investigated by considering some friction. The increase of the longitudinal momentum after averaging over the laser cycle is \cite{Meyer-ter-Vehn1999}
\begin{eqnarray}
	\left\langle \frac{dp_x}{dt}\right\rangle\approx \frac{\nu_\perp a_0^2}{2[1+\left(\gamma\nu_\perp/\omega_0\right)^2]}-\nu_\parallel\left\langle p_x\right\rangle
\end{eqnarray}
where $\nu_\parallel$ and $\nu_\perp$ denote the friction constants along the $x$- and $y$-direction. Since only friction in the $y$-direction has a strong effect on the acceleration, we can assume $\nu_\parallel=0$ for simplicity. For those forward-moving electrons with $p_x\gg p_y $, we can rewrite the above equation as
\begin{eqnarray}
	\left\langle \frac{d\gamma}{dt}\right\rangle\approx \frac{\nu_\perp a_0^2}{2m_ec}\frac{1}{1+\left(\gamma\nu_\perp/\omega_0\right)^2}
\end{eqnarray}
Then one can easily see that the electron energy increases with the time. In the case of $1\gg\left(\gamma\nu_\perp/\omega_0\right)^2$ (small friction), we have $\left\langle\gamma\right\rangle\propto t$, while in the case of $1\ll\left(\gamma\nu_\perp/\omega_0\right)^2$ (large friction), we have $\left\langle\gamma\right\rangle\propto t^{1/3}$. This is broadly consistent with the numerical results shown in Ref. \cite{Meyer-ter-Vehn1999} where they found the effective temperature scales with the interaction time $t^q$ with $q\approx0.5$--1.0. 

In our scheme, when electrons jump across filaments, they experience transverse friction coming from the self-generated electromagnetic fields surrounding the filaments. Therefore, the above simple model can be used to describe the main physical process in our scheme. However, considering the evolution of the fields over multi-ps and also the relativistic effects, it is impossible to quantitatively give an exact derivation of the exponent $q$. Therefore, we have to resort to PIC simulations.  

\begin{figure}
	\includegraphics[width=8.6cm]{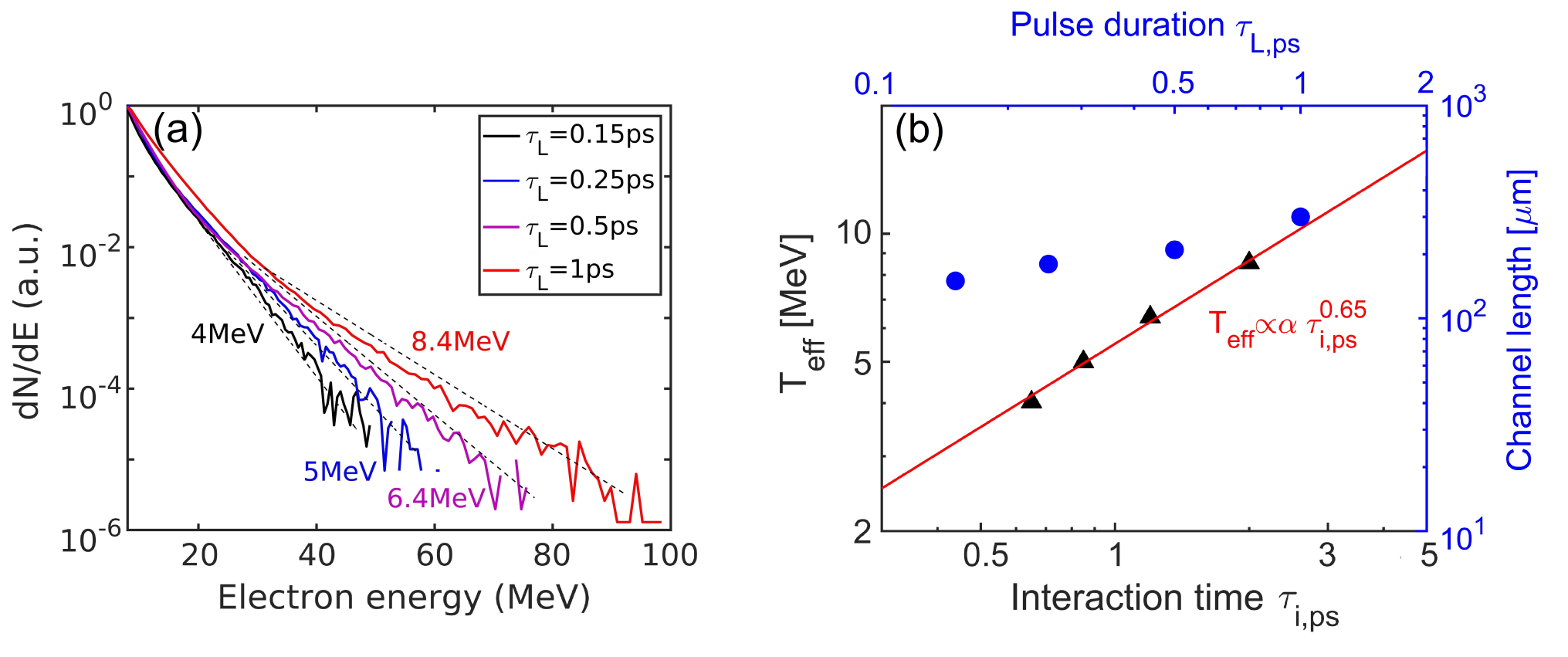}
	\caption{
		\label{fig:fig7}
		(a) Electron energy spectra from different pulse durations. (b) $T_{\rm eff}$ vs  $\tau_{i,ps}$, where the red line represents the best fit and the blue circles show the channel length of each case.  
		Here the energy spectra are integrated over the solid angles, $n_e=0.75n_c$ and $I_0=4.74\times10^{19}\;{\rm W/cm^2}$.}
\end{figure}

To obtain the electron energy scaling of our scheme, we performed further 3D PIC simulations by varying $\tau_L$. 
The final energy spectrum of electrons from each simulation is shown in Fig. \ref{fig:fig7}(a). 
The scaling can be given as [see Fig. \ref{fig:fig7}(b)]
\begin{eqnarray}
	T_{\rm eff}\sim\alpha(I/I_{18})^{0.5}(\tau_{i,ps})^{0.65},
	\label{eq:scaling}
\end{eqnarray}
where the coefficient $\alpha\approx0.8$ and $\tau_{i,ps}$, normalized to 1 ps, is the interaction time. The exponent of 0.65 is within the range predicted by the simple analytical model \cite{Meyer-ter-Vehn1999}. 
Here we determine $\tau_{i}=\tau_L+l_{ch}/c$, where $l_{ch}$ is the channel length observed in the simulations, as specified by the blue circles in Fig. \ref{fig:fig7}(b).

\subsection{Triggering condition}
In the CFSA, the long-surviving multiple filaments are crucial. Conditions for the CFSA as the predominant acceleration mechanism are given by 
\begin{eqnarray}
	d_L\frac{\omega_{rp}}{c}&\sim& 15, \label{eq:r}\\
	2t_i\geq\tau_{L}&\gg& 2\pi/k_fc,  \label{eq:t}
\end{eqnarray}
where  $\omega_{rp}=\omega_p/\sqrt{\gamma}$ is the relativistically-corrected plasma frequency and $t_i=2^{1/4}\sqrt{m_i/Za_0m_e}r_L/c$ is the time for ions to move across the laser spot size. Eq. (\ref{eq:r}) describes the threshold of the filamentation instability \cite{Borisov1995,Huang2015} and Eq. (\ref{eq:t}) determines that many filaments can survive when the laser peak arrives \cite{Gibbon2005}, while electrons have time to jump across multiple filaments. For the parameters considered here, we estimate that $d_L\sim10\lambda$ and $t_i\approx4$ ps. 
Considering the status of kJ, ps laser facilities, the CFSA should always be the predominant acceleration mechanism in LPI with long NCD plasmas unless $\tau_{L}\gg t_i$.

\begin{figure}
	\includegraphics[width=8.6cm]{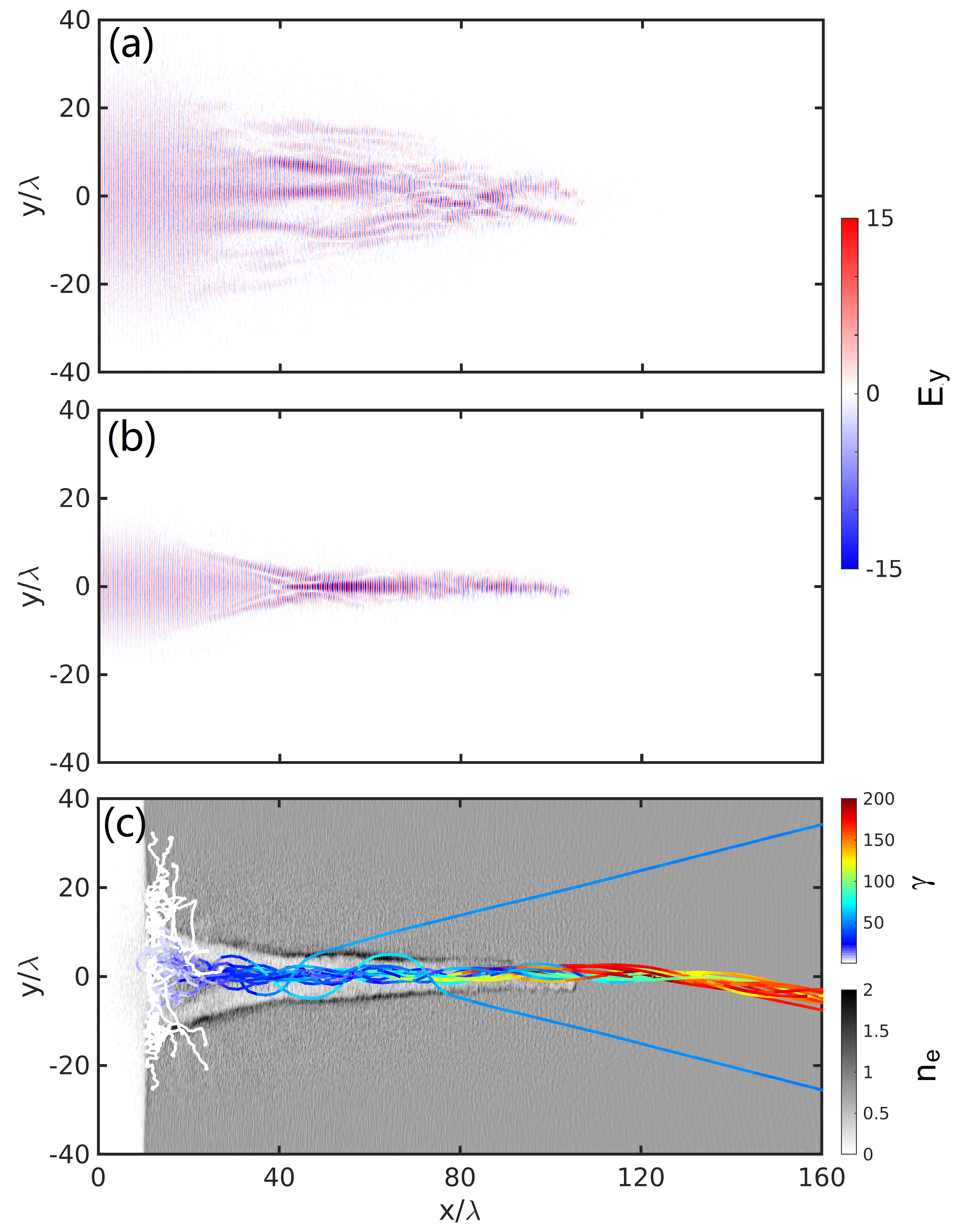} \caption{(a), (b) Distributions of the laser electric field $E_y$ at $t=0.117$ ps for the cases with laser focal spot $d_L=23.6\lambda$ and $11.8\lambda$, respectively. (c) Electron trajectories (color lines) and density distribution for the case with $d_L=11.8\lambda$, showing that with a small focal spot, most of energetic electrons are confined in a single channel and accelerated via the DLA mechanism. In (c), the solid lines show trajectories of the selected electrons with their relativistic $\gamma$-factor color-coded. 	
	}
	\label{fig:fig8} 
\end{figure}

According to Eq. (\ref{eq:r}), to suppress the filamentation instability, it requires $d_L\sim12.3\lambda$ where we assume $\gamma=20$ and $n_e=0.75n_c$ based on the parameters used in Fig. \ref{fig:fig5}. As a demonstration, we show 3D PIC simulations with laser focal spot sizes of larger ($d_L=23.6\lambda$) and smaller ($d_L=11.8\lambda$) than the threshold in Figs. \ref{fig:fig8}(a) and \ref{fig:fig8}(b), respectively.  
One can clearly see that when the focal spot size is smaller than the threshold, a single channel is formed [Figs. \ref{fig:fig8}(b) and (c)], while when it is larger, multiple filaments appear. In Fig. \ref{fig:fig8}(c), we show the trajectories of electrons and electron density distribution for the single channel case, where it is evident that most of energetic electrons are confined inside the channel and undergo betatron oscillation. This is a typical phenomena of DLA.

\begin{figure}
	\includegraphics[width=8.6cm]{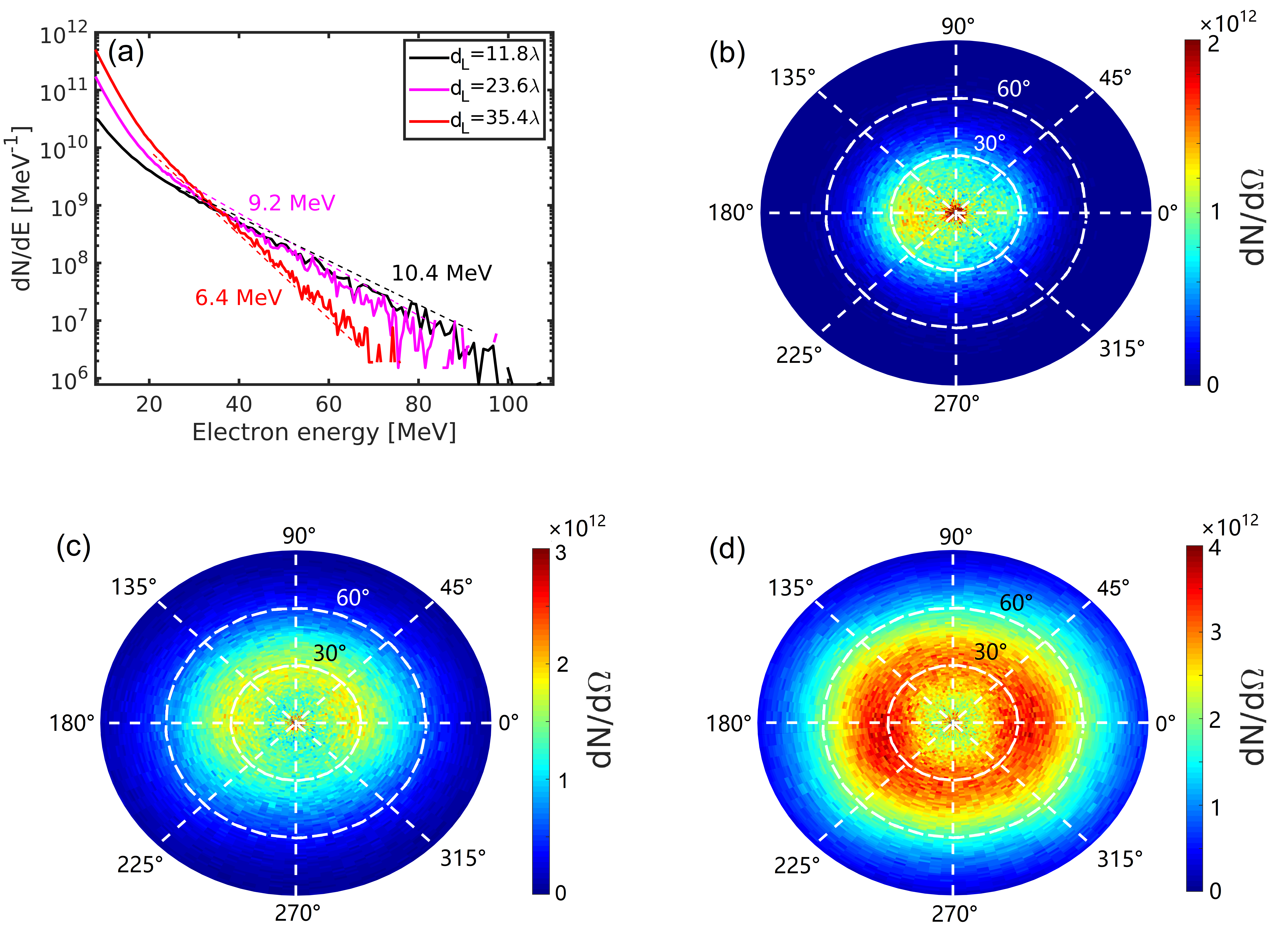} \caption{
		(a) Electron energy spectra registered at the boundaries for the cases with focal spot size $d_L=11.8\lambda$ (black), 23.6$\lambda$ (magenta) and 35.4$\lambda$ (red), while (b)-(d) correspondingly show the angular distribution of high-energy electrons. Note that here the energy spectra are integrated over the solid angles, $n_e=0.75n_c$, $I_0=4.74\times10^{19}\,{\rm W/cm^2}$ and $\tau_L=500$ fs. 	
	}
	\label{fig:fig9} 
\end{figure}

\section{SUMMARY AND DISCUSSION}
As we discussed in the above section, with the increase of the laser focal spot size, the electron acceleration may transit from the DLA to the CFSA due to the arise of the filamentation instability. To show the difference of features of the  electron beams obtained from these two mechanisms, in Fig. \ref{fig:fig9}(a), the energy spectra of electrons from the cases with focal spot size $d_L=11.8\lambda$, 23.6$\lambda$ and 35.4$\lambda$ are depicted by the solid black, magenta and red lines, respectively, while the angular distributions are correspondingly shown in Fig. \ref{fig:fig9}(b)-(d). 

One can see that with a narrow pulse where the DLA predominates, the obtained electron beam has a higher electron temperature and maximum electron energy compared to the cases from wider pulses where the CFSA predominates. This is because in DLA, electrons can stay in the acceleration phase for longer time. Moreover, as shown in Fig. \ref{fig:fig9}(b), the electron beam from DLA is collimated to the forward direction while in the CFSA [see Figs. \ref{fig:fig9}(c) and \ref{fig:fig9}(d)], the peak brightness deviates from the forward direction due to the random scattering. This feature can be used in experiment to distinguish the CFSA from the DLA.

We mention that the electron flux from the CFSA is much higher than that obtained from the DLA. For the case with $d_L=35.4\lambda$, the number of electrons with energy larger than 2.5 MeV reaches 8.77$\times10^{13}$ ($14\mu$C, registered at the boundaries), which is about 16.5 times higher than that with $d_L=11.8\lambda$, though the laser energy is only about 9 times higher. This demonstrates that the CFSA is more suitable for producing high-flux electrons compared to the DLA or other known mechanisms, since in the DLA, only small part of electrons can reach the resonant condition to be accelerated to high energies. Moreover, the divergence in our scheme is comparable to or even smaller than that observed in electron-induced fast ignition \cite{Kemp2014,Robinson2014,Jarrott2016} where the electron divergence half-angle is always greater than 30°, and possibly exceeding 50°--60° \cite{Robinson2014}. To guide the electron transport, a cone target can be used just like that in fast ignition. On the other hand, neutron sources are usually isotropic \cite{Guenther2022}, and therefore the divergence of the electrons/protons is not that important, especially for the bulk target regime \cite{Willingale2011}.

Therefore, our work would significantly facilitate applications in HED science, particle and radiation sources, fusion energy, nuclear science, etc., where  
the electron flux is the key parameter, rather than the temperature or maximum energy. 
For example, towards the application of electron-induced fast ignition fusion \cite{Kemp2014,Robinson2014}, the required electron temperature is about 1--3 MeV to ensure that electrons can reach the dense central core and be stopped, but the required flux is extremely high (i.e., $\sim4\times10^{16}$ electrons within 10--20 ps) \cite{Robinson2014}. Though this flux is higher than our present results, by considering appropriate laser parameters \cite{Kemp2014}, it might be possible to approach the requirements via the CFSA especially due to the high conversion efficiency in our scheme.  
In the applications of high-flux laser-based neutron sources \cite{Willingale2011,Brown2018,Guenther2022} or proton-induced fast ignition fusion \cite{Roth2001}, only moderate ion energies are required while the ion flux should be huge. In $\rm ^7Li(p,n)^7Be$ reaction, the proton energy of the peak of the cross section is only several MeV \cite{Brown2018}. To produce such proton beams, the required electron temperature should not be high, where several MeV is sufficient \cite{Macchi2013,Qiao2019,Shen2021b}. 

Furthermore, high-flux electrons can also be used to produce high-flux x-rays via Bremsstrahlung \cite{Koch1959,Hollinger2017} or betatron radiation \cite{Shen2021a}. This is important for developing x-ray based diagnostic techniques to probe the implosion dynamics in ICF and shock waves in related HED science, where high-flux is a crucial parameter to overcome the self emission of the plasma and achieve good statistics \cite{Albert2016}. On the other hand, generation of high-flux superponderomotive electrons is essential for optimizing target normal sheath acceleration (TNSA) proton sources which  have become a powerful tool for probing and characterizing HED plasmas in recent years \cite{Simpson2021b,Borghesi2001}.

Though our discussion is focused on the electron acceleration in long homogeneous NCD plasmas, it should also be applicable for interactions of kJ, ps lasers with solid targets where a long-scale length preplasma appears. The preplasma can be induced by either the prepulse \cite{Simpson2021,Raffestin2021} or the long-time interaction \cite{Kemp2012}. 
The formation of multifilaments in long-scale length preplasmas has been observed  \cite{Tanaka2000,Pukhov2003,Pugachev2016}. Therefore, we deduce that the CFSA can also contribute to generation of superponderomotive electrons in interactions of kJ, ps lasers with solid targets \cite{Williams2020}. Moreover, we mention that due to the long-time evolution of plasma and field structures, other mechanisms may also contribute to the electron acceleration \cite{Sorokovikova2016,Iwata2018,Kemp2020,Pukhov1999}, but the CFSA may manifest itself in the angular distribution of electrons [see Fig. \ref{fig:fig3}(b)] if it becomes the predominant mechanism.

In conclusion, we have performed the first systematic study on interactions of realistic kJ, ps lasers with submillimeter NCD plasmas. We demonstrate that due to multifilaments formation, copious electrons jump across multichannels and obtain superponderomotive energies from the stochastic acceleration. We find the electron temperature grows with the pulse duration.  
Our work provides an attractive approach for producing high-flux electron beams and secondary sources, including ions \cite{Roth2001}, neutrons \cite{Brown2018}, x-rays \cite{Koch1959,Hollinger2017}, etc., which can be further used in important applications from diagnostic techniques to fusion energy.

\section*{Acknowledgements}

This work is supported by the DFG (project PU 213/9). The authors gratefully acknowledge the Gauss Centre for
Supercomputing e.V. for funding this project  by providing computing time
through the John von Neumann Institute for Computing (NIC) on the GCS Supercomputer JUWELS at J\"ulich Supercomputing Centre (JSC). The research of N.E.A. was supported by The Ministry of Science and Higher Education of the
Russian Federation (Agreement with Joint Institute for High Temperatures RAS No 075-15-
2020-785 dated September 23, 2020). X.F.S. gratefully acknowledges support by the Alexander
von Humboldt Foundation, as well as helpful discussions with
L. Reichwein at HHU.\bigskip{}

\end{document}